# Switching current distributions and subgap structures of underdoped (Hg,Re)Ba$_2$Ca$_2$Cu$_3$O$_{8+\delta}$ intrinsic Josephson junctions

S. Ueda,[1] T. Yamaguchi,[2] Y. Kubo,[2] S. Tsuda,[2] J. Shimoyama,[3] K. Kishio,[3] and Y. Takano[2]

[1]*Department of Applied Physics, Tokyo University of Agriculture and Technology, 2-24-16 Nakacho, Koganei, Tokyo 184-8588, Japan*

[2]*National Institute for Materials Science (NIMS), 1-2-1 Sengen, Tsukuba 305-0047, Japan*

[3]*Department of Applied Chemistry, University of Tokyo, 7-3-1 Hongo, Bunkyo-ku, Tokyo, 113-8656, Japan*

(Dated: November 8, 2009)

We have investigated the intrinsic Josephson properties in slightly underdoped (Hg,Re)Ba$_2$Ca$_2$Cu$_3$O$_y$ [Hg(Re)1223] intrinsic Josephson junctions (IJJs) with dimension of 1.0 x 1.5 x 0.11 μm$^3$. The current-voltage characteristics of the IJJs exhibit clear multiple branches with subgap structures similar to those of other cuprate superconductors. The switching current distributions, $P(I)$, from the zero-voltage to a nonzero-voltage state in the current-biased IJJs agree well with the theoretical curves of the thermally assisted escape model at temperatures above ≈ 5 K. The plasma frequency, $f_p$, of the IJJs is estimated to be 1.3 THz from the fluctuation-free critical current density of 2.0 x 10$^5$ A/cm$^2$, which is one of the highest among cuprate superconductors, reflecting the high $T_c$ and a relatively low anisotropy of the Re doped Hg system. The $P(I)$ gradually becomes independent of temperature below ≈ 5 K, which suggests a crossover of the escape process from thermal activation to quantum tunneling at such a high temperature.

## I. INTRODUCTION

The intrinsic Josephson junctions (IJJs), which consist of alternately stacked superconducting and barrier layers in cuprate superconductors, are one of the promising candidates for quantum device application such as qubits. In the last few years, the observation of macroscopic quantum tunneling (MQT) and energy level quantization has been successfully demonstrated in current-biased IJJs in Bi$_2$Sr$_2$CaCu$_2$O$_y$ (Bi2212).[1-6] The crossover temperature between thermal and quantum escape of Josephson phase is relatively high up to ≈ 1 K for Bi2212 IJJs because of the high Josephson plasma frequency of about 100 – 200 GHz. The crossover temperature is proportional to the plasma frequency,[7] which is proportional to the square root of the critical current density, $J_c$, and therefore IJJs with a large $J_c$ are advantageous to observe MQT at high temperatures.

Among high $T_c$ superconductors, a series of the mercury-based superconductors HgBa$_2$Ca$_{n-1}$Cu$_n$O$_{2+2n+\delta}$ [Hg12($n$-1)$n$; $n$ = 1, 2, 3,··] has high $T_c$ with the highest $T_c$ of ≈ 135 K for a compound with $n$ = 3. The thickness of barrier layers of this system, ≈ 9.5 Å, is slightly thinner than that of the Bi system, ≈ 12.1 Å, which result in a somewhat lower electromagnetic anisotropy than that of Bi system. The high $T_c$ and the relatively low anisotropy of the mercury-based family bring in a high $J_c$ along $c$-axis; 3.4 x 10$^5$ A/cm$^2$ at 4.2 K for Hg(Re)1234,[8] 1.0-1.4 x 10$^5$ A/cm$^2$ at 10 K for Hg(Re)1212,[9] while about 10$^3$ A/cm$^2$ at 4.2 K for Bi2212. The IJJs of the Hg system are expected to have high potential for observing MQT at high temperatures. However, because of considerable difficulties in the synthesis of high quality single crystals or high-quality epitaxial thin films, a precise understanding of the $I$-$V$ characteristics and measurements of switching dynamics have been lacking. In our previous study, we have successfully grown high quality single crystals of Hg(Re)12($n$-1)$n$ with $n$ = 2, 3, and 4 by the flux method.[10,11] The single crystals are favorable to fabricate IJJs because they are typically plate-like with wide $ab$-plane and well cleavable like Bi2212 single crystals.

Since IJJs in the Hg system have very thin superconducting layers 3.2 x ($n$-1) Å, and have large $J_c$ as mentioned above, they have a very short Josephson penetration depth, $\lambda_J$, calculated to be in the order of several ten nm.[12] Hence, even if the lateral dimension, $L$, of the IJJs is reduced to 1 μm, the junction is considered to be still in a long-junction limit where the phase difference across the junction can vary in space. In this case, the switching dynamics of the junctions may be determined by the configuration of fluxons in the IJJs, as was found in Bi2212 large IJJs with $L$ ≈ 15 μm.[13] This is in contrast to the case of Bi2212 IJJs with $L$ ≈ 1 μm in which electrodynamics could be accurately described neglecting the variation of the phase difference between electrodes.[1-6] $\lambda_J$ of Bi2212 IJJs has been estimated to be submicron order,[12] so that $L$ of the junctions is close to or slightly larger than the $\lambda_J$. In a long Josephson junction made of conventional metal superconductors, it has been reported that single fluxon flow by the current perpendicular to the junction showed a quantum tunneling at low temperatures.[14] This result showed a prospect of designing a vortex qubit utilizing Josephson vortices, and implied that large IJJs may be possible candidates for the Josephson vortex qubits.

Unlike Bi2212 IJJs, which exhibit heavily underdamped characteristics, Re doped Hg systems with a lower anisotropy have more conductive barrier layers,[10,15] so that the enhancement of electrical isolation of barrier layers is needed to observe typical SIS properties like those of Bi2212 IJJs. In this study, we prepared slightly underdoped Hg(Re)1223 IJJs with $T_c$ of 129 K by controlling the oxygen nonstoichiometry to decrease the $c$-axis conductivity. The $I$-$V$ curves of the IJJs exhibit a clear multiple-branch structure, which suggests an interference of Raman optical phonons with ac Josephson current. We have investigated the switching dynamics of the IJJs and the results indicate that a crossover from thermal activation (TA) to MQT occurs at a temperature as high as ≈ 5 K.

## II. EXPERIMENT

Intrinsic Josephson Junctions with dimensions of 1.0 x 1.5 x 0.11 μm$^3$ were fabricated using a focused Ga$^+$ ion beam from a $Hg_{0.84}Re_{0.16}Ba_2Ca_2Cu_3O_{8+\delta}$ single crystal grown by a flux method.[10,11] The details of the fabrication process are described elsewhere.[16] The overlapped length through the $c$-axis, 0.11 μm, corresponds to the total junction number of ≈ 70. The career doping state of the IJJs was tuned by vacuum reduction treatment ($P_{O2}$ < 10$^{-2}$ Pa) to a slightly underdoped region where the $T_c$ was 129 K.

The $I$-$V$ characteristics of IJJs were measured using the four probe method in a $^3$He cryostat with a minimum temperature of 600 mK. In order to reduce the influence of high frequency noise, we mounted the sample in a copper cell and used lossy coaxial cables with the nominal attenuation of 80 dB at 5 GHz. The voltages across the sample and a reference resistance to monitor the current were amplified using battery-operated low noise amplifiers.

The switching current distribution, $P(I)$, from its zero-voltage to a finite voltage state was measured by measuring the time delay between zero crossing of a linear bias current ramp applied to the sample and switching of the junction to the finite voltage state. The samples were biased using a constant ramp rate, $dI/dt$ = 0.1 - 0.3 A/s, with a repetition rate of 20–30 Hz using a function generator, Agilent 33220A. After detecting a switching event by a voltage threshold of 40 μV, the current was switched to zero within less than 10 μs and kept zero for more than 130 ms in order to avoid heating effects. The measurement was performed using a universal time interval analyzer, YOKOGAWA TA320, with time resolution of 100 ps. The switching current at each temperature was recorded 7000 - 10000 times repeatedly.

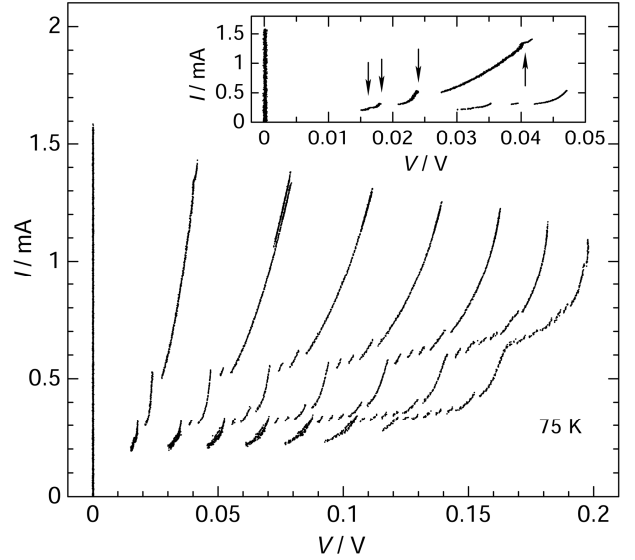

FIG. 1. Multi-branch structure of the Hg(Re)1223 IJJs at 75 K. The inset shows the first branch with arrows indicating the current jumps.

## III. RESULTS AND DISCUSSIONS

Figure 1 shows a plot of the typical multi-branch structure in the underdoped Hg(Re)1223 IJJs at 75 K. The magnitude of the voltage jump to the first branch is ≈ 42 mV, which corresponds to the large superconducting gap of Hg system and is larger than 20 – 30 mV reported for Bi and Tl systems. The multi-branch structure is observed up to around 115 K.

In the $I$-$V$ characteristics, clear subgap structures are seen in each quasi particle branch. On the $n$th branch, there are $n$ sub-branches in each region. The inset of fig. 1 shows the enlarged first branch with arrows indicating the position of current peaks. The current peaks with voltage jumps on the first branch are seen at 18.2 and 24.0 mV, and weak current peaks are also seen at 16.7 and 40.5 mV. These characteristic voltages are relatively higher than those of Bi2212 and Tl2223 IJJs, which are lower than 10 mV.[17,18] As common features of subgap structures in cuprate IJJs, the characteristic voltages do not depend on temperature, junction geometry or magnetic fields. Figure 2 is the $I$-$V$ curves of the Hg(Re)1223 IJJs at 4.23 K and 90 K, and it shows that the voltages for the subgap structures are not changed by temperature. We also verified that the voltage positions did neither depend on junction geometry, nor magnetic fields up to 1 Tesla below $0.75T_c$.

Several mechanisms have been introduced to explain the appearance of subgap structures.[19,20]
One is the resonant coupling mechanism between infrared active optical $c$-axis phonons and oscillating Josephson currents. Another is the phonon assisted tunneling mechanism due to the interference of Raman-active optical phonons with ac Josephson currents in the junctions. Ponomarev *et al* showed that the fine structures at subgap voltages in the $dI/dV$ curves of Bi2212 break junctions were in good agreement with the Raman scattering spectra of the phonon.[21] According to either explanation, current peaks appear when the voltage satisfies the equation: $2eV = \hbar\omega$, where $\omega$ is infrared active optical c-axis phonon frequency for the former, and Raman-active optical phonon frequency for the latter.

Table I. Voltages of subgap structures and corresponding frequencies of the Hg(Re)1223 IJJs compared with the Raman-active optical phonons of Hg(Re)1223 and calculated infrared active phonons for Tl1223. $O_{IP}$, $O_{OP}$, and $O_{ap}$ indicate the oxygen at inner $CuO_2$ plane, at outer $CuO_2$ plane, and the apex oxygen of $ReO_6$ octahedra, respectively.

| Present work (Hg(Re)1223) | | Raman-optical phonons ( Hg(Re)1223) | | IR optical c-axis phonons (Tl1223) | |
|---|---|---|---|---|---|
| $V$ [mV] | $\omega$ [cm$^{-1}$] | $\omega$ [cm$^{-1}$] | Assignment | $\omega$ [cm$^{-1}$] | Assignment |
| 16.7 | 269 | 265 | $O_{OP}$[a] | 245 | Ca, $O_{IP}$, $O_{OP}$[b] |
| 18.2 | 293 | 285 | Ca[c] | | |
| | | 300 | Ca[c] | | |
| 24.0 | 387 | 400 | $O_{OP}$[a] | 372 | $O_{IP}$, $O_{OP}$[b] |
| 40.5 | 653 | 640 | $O_{OP}$[c] and $O_{ap}$[c] | | |

[a]Measurements of a Hg1223 single crystal at 10 K. (Ref. 22)
[b]Ref. 25
[c]Measurements of a Hg1223 single crystal at room temperature. (Ref. 23)
[d]Measurements of Re doped Hg1223 polycrystalline at room temperature. (Ref. 24)

Table I summarized the characteristic voltages and the corresponding frequencies of subgap structures in the Hg(Re)1223 IJJs in comparison with the Raman-active optical phonons of Hg(Re)1223 and infrared active phonons calculated for Tl1223.[22-25] We used the data for Tl1223 for infrared active phonons because that of Hg(Re)1223 has not been investigated yet. In the table, the selected phonons, close to the frequencies of present work, from all phonon modes are shown. It seems that the characteristic frequencies for the subgap structures correspond better with the frequencies of Raman optical phonons than with those of infrared active phonons, which implies that the appearance of subgap structures is attributed to an interference of Raman optical phonons with ac Josephson current. For a more precise understanding of the subgap structures, high resolution low-energy infrared and Raman spectra for Hg(Re)1223 are needed.

The upward switching current, $I_c$, of the IJJs is typically around 2.86 mA at 4.2 K as shown in fig. 2. The value of McCumber parameter, $\beta_c$ ($\approx (4I_c/\pi I_r)$) is estimated to be ≈1300 from the ratio $I_c / I_r \approx 32$, where $I_r$ is return current at which the quasi-particle branch jumps back to the zero voltage. The $J_c$ value of Hg(Re)1223 IJJs at 4.2 K is as high as $1.9 \times 10^5$ A/cm$^2$, which is two orders of magnitude larger than that of Bi2212. Therefore, the Hg(Re)1223 IJJs have a very short Josephson penetration length (= 33 nm).

The inset of fig. 2 shows the temperature dependence of $I_c$ derived from I-V curves at various temperatures. The dashed line indicates the theoretical curve based on the Ambegaokar–Baratoff (AB) relation for the SIS-type Josephson junction.[26] The $I_c$–$T$ curve largely deviates from the AB predictions in the middle temperature range, and furthermore, multiple-valued $I_c$ is observed at around 40 K. Such deviations and the multiple-valued $I_c$ have also been observed by Mros et al in large Bi2212 IJJs,[13] and they attributed it to the presence of Josephson fluxon in the junctions in a multiplicity of different configurations. Each fluxon configuration results in a different value of apparent $I_c$. In Hg(Re)1223, the $\lambda_J \approx 33$ nm is very shorter than the lateral dimensions of the junction, ≈ 1 μm; hence the anomaly in $I_c$-$T$ curves in Hg(Re)1223 could be considered as the same behavior of Bi2212 large junction.

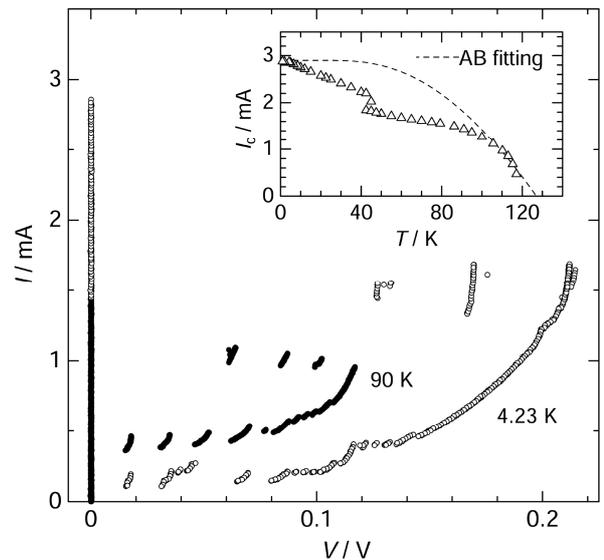

FIG. 2. Multi-branch structure of the Hg(Re)1223 IJJs at 4.23 K and 90 K are shown. The inset shows temperature dependence of $I_c$ of the IJJs with dashed line representing the Ambegaokar-Baratoff (AB) theory.

The similar deviations in $I_c$-$T$ curves have also been observed in YBa$_2$Cu$_3$O$_y$ IJJs which have barrier layers with a high conductivity and have a small $\beta_c \approx 10$.[27] Such IJJs are close to the SNS-type junction rather than the SIS-type junction. Although the value of $\beta_c$ of the Hg(Re)1223 IJJs is much larger than that of Y123 IJJs, there is a possibility that the Hg(Re)1223 IJJs may not be a simple SIS type because their barrier layers contain a conductive ReO$_6$ structure.

Now we turn to the switching current properties in the current-biased IJJs of Hg(Re)1223. In a large area Josephson junction, the phase difference between electrodes may vary in space, which is in contrast with a small Josephson junction where the spatial variation of the phase is negligible. The small Josephson junction model which has often been used for Bi2212 IJJs with dimensions of several micron can not explain the behavior of $P(I)$ of our Hg(Re)1223 IJJs probably because its $\lambda_J$ is extremely smaller than $L$, so that we analyze our data using a long junction theory proposed in Ref. 13. As is well known, the dynamics of a current-biased Josephson junction can be described by an equation of the damped motion of a particle in a tilted-washboard potential. The escape rate $\Gamma_T$ from the $V = 0$ state of a Josephson junction in the low damping regime can be analyzed in terms of activation model, which is given by:[28,29]

$$\Gamma_T = \frac{\omega_0}{2\pi} \exp\left(-\frac{U_0}{k_B T}\right) \quad (1)$$

where $U_0$ is the barrier energy of the meta stable state and $\omega_0$ is the bias depending plasma frequency. In a long junction, $U_0$ is the energy required to create a fluxon in a junction given by $U_0 = 8\Phi_0 I_0 \lambda_J/(2\pi L)$ where $\Phi_0$ ($\equiv h/(2e)$) is the magnetic flux quantum and the $I_0$ is the fluctuation-free critical current. The energy barrier for long junction with taking sample geometry and the bias current dependence into consideration is

$$U_0 = 8V_0 \frac{\Phi_0 I_0 \lambda_J}{2\pi L}\left(1 - \frac{I}{I_0}\right)^{3/2} \quad (2)$$

where $V_0$ is a dimensionless constant depending on the sample geometry. At temperatures below a crossover temperature, $T^* \approx \hbar\omega_0/2\pi k_B$, the escape is dominated by quantum tunneling through the barrier. In the MQT process, the escape rate $\Gamma_{MQT}$ is given by [30]

$$\Gamma_{MQT} = 12\omega_0\left(\frac{3U_0}{2\pi\hbar\omega_0}\right)^{1/2} \exp\left(-\frac{36U_0}{5\hbar\omega_0}\right) \quad (3)$$

Incidentally, the bias depending plasma frequency of a small Josephson junction is

$$\omega_0 = \omega_P\left[1 - (I/I_0)^2\right]^{1/4} \quad (4)$$

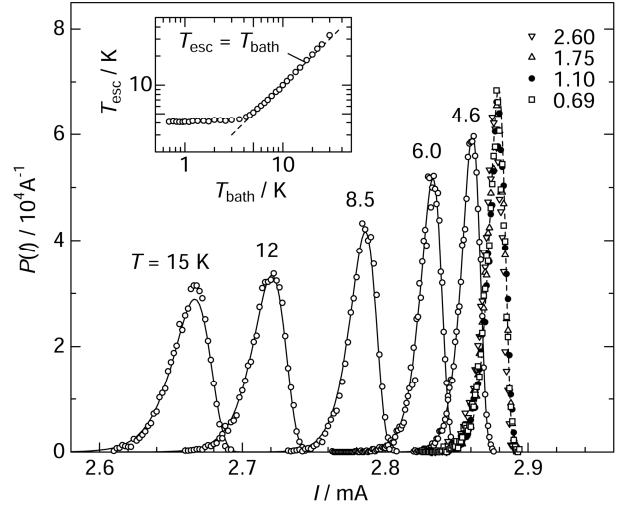

FIG. 3. Switching current distributions, $P(I)$, of the IJJs measured at 0.69 – 15 K. The dots are the experimental data and the solid and dotted lines show the fits of the equation (1) and (3). The inset shows the escape temperature vs bath temperature. The dashed line indicates $T_{esc} = T_{bath}$.

where $\omega_p = 2\pi I_0/\Phi_0 C$ is the zero-bias plasma frequency and $C$ is the junction capacitance. We use the short junction value of $\omega_0$ for our junction as in Ref. 13, considering that it does not affect significantly the results in thermally activated region. The bias dependence of the characteristic attempt frequency of a large junction is the same as that of a short junction limit for $I \to I_0$,[31] and $\Gamma$ depends linearly on $\omega_0$ shown in equation (1), which indicates that a shift in $\omega_0$ has a much smaller influence on the escape rate than a shift in the energy barrier. On the other hand, $\Gamma$ depends exponentially on $\omega_0$ in MQT region, so that we can not adopt the attempt frequency of the small junctions to the equation (3).

Figure 3 shows a plot of switching current distribution, $P(I)$, in the Hg(Re)1223 IJJs as a function of the bias current $I$, measured at 0.69 – 15 K. The dots represent the experimental data, and the solid lines represent the theoretical fitting based on equation (1) using escape temperature, $T_{esc}$, as a fitting parameter. The distribution of $P(I)$ tends to be sharper as the temperature is lowered, and the data agree well with the prediction of TA theory.

The fitting of $P(I)$ gives the zero-noise critical current $I_0 = 3.021$ mA using $dI/dt = 0.1067$ A/s and $C = 144$ fF. The junction capacitance is calculated assuming the dielectric constant of the barrier $\varepsilon_r = 10$. It is notable that the $I_0$ is significantly large compared to other cuprates, which results in a large plasma frequency, $f_p$ ($\equiv \omega_p/2\pi$) = 1.3 THz. From 2.6 K to 0.69 K, $P(I)$ no longer depends on temperature. The dashed line represents equation (3) with an appropriate attempt frequency used to fit the experimental data. We made sure that the measurement-system noise is not the main reason for the $P(I)$ saturation by measuring other IJJs with much lower $J_c$. In the inset of fig. 3, the escape temperature $T_{esc}$ for the first switching is plotted as the bath temperature, $T_{bath}$. We find good agreement between $T_{esc}$ and $T_{bath}$ for temperatures above 5 K, and then $T_{esc}$ keeps a nearly constant value below the temperature. The temperature independent switching suggests the possibility of MQT. We have measured another two samples and observed similar trends at a few Kelvin.[32]

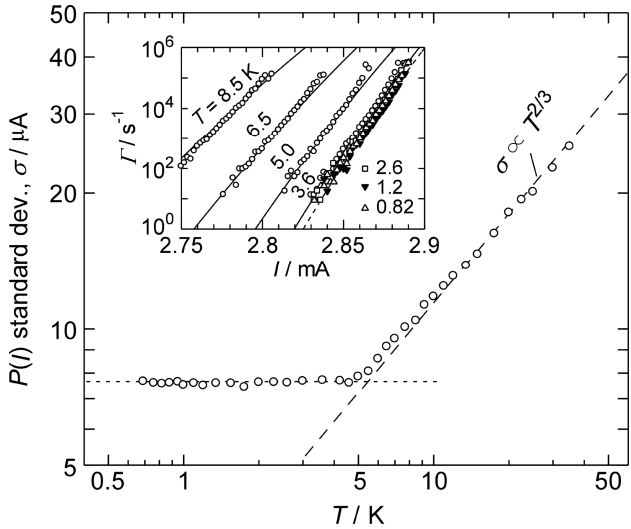

FIG. 4. The $P(I)$ standard deviation $\sigma = (\langle I^2\rangle - \langle I\rangle^2)^{1/2}$ as a function of temperature between 0.69 K and 35 K. The dashed and dotted lines indicate $\sigma \propto T^{2/3}$ and a linear fit to the data points below 5 K. The inset shows escape rate, $\Gamma(I)$, of the IJJs at 0.82 – 8.5 K compared with the theoretical escape rate. The dots are the experimental data and the solid and dotted lines show the fits of the equation (1) and (3).

Figure 4 shows the width of $P(I)$, $\sigma (= (\langle I^2\rangle - \langle I\rangle^2)^{1/2})$, as a function of temperature. The $\sigma$, which is a measure of the strength of thermal fluctuations, is approximately proportional to $T^{2/3}$ in a range of 5 – 35 K, which indicates that the escape process is dominated by thermal activation in this temperature regime. The saturation of $\sigma$ below approximately 5 K indicates the crossover from TA to MQT regime. The theoretically predicted crossover temperature $T^* \approx \hbar\omega_0/2\pi k_B$ is calculated to be 5.4 K using $\omega_0$ derived by equation (4) with $f_p = 1.3$ THz and $I/I_0 = 95\%$. The temperature agrees quantitatively with the experimental value. Because of the quite high $f_p$ of the Hg(Re)1223 IJJs, almost ten times higher than that of Bi2212, about 100 – 200 GHz, the MQT can occur at temperatures several to ten times higher than that of Bi2212.[1-6] The inset of fig. 4 shows escape rate, $\Gamma$, calculated from the experimental data with theoretical escape rates for comparison. The dots are the experimental data and the solid lines and dotted line shows the theoretical fitting represent the equation (1) and (3). $\Gamma$ gradually becomes independent of temperature below 5 K and almost complete overlap is verified below 2.6 K. Our results indicate potential advantage of Hg system among high $T_c$ superconductors applicable to quantum device applications.

## IV. CONCLUSIONS

We have investigated the intrinsic Josephson properties in Hg(Re)1223, and observed clear subgap structures similar to those of other cuprate superconductors. The characteristic frequencies for the subgap structures are in rough correspondence with the frequencies of Raman optical phonons in Hg1223. The switching current distributions of Hg(Re)1223 IJJs agree well with the prediction of TA theory down to approximately 5 K and then gradually becomes independent of temperature, which is attributed to the crossover of escape process from the thermally activated regime to the quantum regime. The crossover temperature is consistent with the theoretically predicted crossover temperature estimated from the plasma frequency of the IJJs, $f_p = 1.3$ THz. The IJJs in a Re doped Hg system exhibit good potential for designing qubits at high temperatures of several Kelvin.